# Exploration of Multi-Scale Image Fusion Systems in Intelligent Medical Image Analysis

Yuxiang Hu[1], Haowei Yang[2], Ting Xu[3], Shuyao He[4], Jiajie Yuan[5], Haozhang Deng[6]

[1]Johns Hopkins University,USA,yhu102@jhu.edu

[2]University of Houston,USA,yanghaowei09@gmail.com

[3]University of Massachusetts Boston,USA,ting.xu001@umb.edu

[4]Northeastern University,USA,he.shuyao@northeastern.edu

[5]Brandeis University,USA,jiajieyuan@brandeis.edu

[6]Northwestern University,USA,haozhangdeng2021@u.northwestern.edu

*Abstract*—The diagnosis of brain cancer relies heavily on medical imaging techniques, with MRI being the most commonly used. It is necessary to perform automatic segmentation of brain tumors on MRI images. This project intends to build an MRI algorithm based on U-Net. The residual network and the module used to enhance the context information are combined, and the void space convolution pooling pyramid is added to the network for processing. The brain glioma MRI image dataset provided by cancer imaging archives was experimentally verified. A multi-scale segmentation method based on a weighted least squares filter was used to complete the 3D reconstruction of brain tumors. Thus, the accuracy of three-dimensional reconstruction is further improved. Experiments show that the local texture features obtained by the proposed algorithm are similar to those obtained by laser scanning. The algorithm is improved by using the U-Net method and an accuracy of 0.9851 is obtained. This approach significantly enhances the precision of image segmentation and boosts the efficiency of image classification.

## I. Introduction

In the past, the edge extraction of brain tumor images was mostly artificial, which was complicated and easy to make mistakes [1]. Medical image segmentation is an important artificial intelligence technology[2-4], which is expected to achieve quantitative determination of brain tumor characteristics in clinical scenes and is currently a hot topic in medical image segmentation research.

Deep learning stems from the study of neural networks, which are abstract computational patterns built to mimic the thinking activity of neurons in the brain. Deep learning approaches have produced impressive outcomes across a range of machine learning tasks, such as geography prediction[5-6], image analysis [7-10], and natural language processing [11-14]. Some researchers have proposed using deep neural networks to block a photo and then learn from a large deep network based on the mid-point markers of each photo to improve segmentation and recognition efficiency [15-18]. Some researchers have established a hierarchical learning model, which greatly improves the efficiency of image segmentation [19]. In recent times, certain researchers have championed multiscale convolutional algorithms, distinguishing them from traditional neural networks by obviating the necessity for single-scale image convolution[20]. Some studies have proposed a medical image processing method U-Net [21]based on a global convolutional network, which can effectively combine high-level semantics with shallow features to improve the accuracy of diagnosis. This project intends to construct a brain tumor boundary extraction algorithm based on U-Net and realize the segmentation of brain glioma MRI by introducing residual networks, models with enhanced context, and a pooled pyramid model based on black cavity space convolution. In this way, accurate segmentation of brain glioma magnetic resonance imaging can be achieved.

(1) By combining Resnet residual mapping mode and U-Net drop sampling mode, as the backbone network of image processing contraction path, the degradation phenomenon of gradient explosion and gradient disappearance in network training is effectively solved, so as to realize the transmission of shallow feature information and the direct integration of the upper network. (2) Using the scene enhancement model, the model is sampled four times to improve the intrinsic characteristics of each pixel. This algorithm not only increases the perceptual domain, but also reduces the computation. (3) The convolution of the hole and the cone of the pool are introduced in this method to achieve multi-scale fusion.

## II. SEGMENTATION MODEL OF BRAIN TUMOR

### A. Overall design of the model

The entire network is divided into compression channels and extended channels (Figure 1 is quoted in a multi-scale feature fusion network for detection). The algorithm also transforms it by the residual and then transfers the extracted high-level characteristic information to the underlying neural network in a "jump" way. A background enhanced model is introduced, and the convolution function is introduced in between. CEBlock plays the role of enhancing background information, and ASPs block performs non-uniform spatial convolution through non-uniform sampling rate, so as to realize multi-scale information fusion [22]. After the CE block and ASPP block are used, the detection of multiple scale objects can be strengthened effectively, and the accuracy of segmentation is obviously improved. Secondly, Decoder block is used to extract the features in the expansion channel[23]. This project intends to use the jump method to combine the feature information obtained after decoding at each level with the feature information extracted from the corresponding drop sampling on the compression path. Then the decoding operation of the next level is carried out to realize the gradual restoration of the feature mapping of each level.

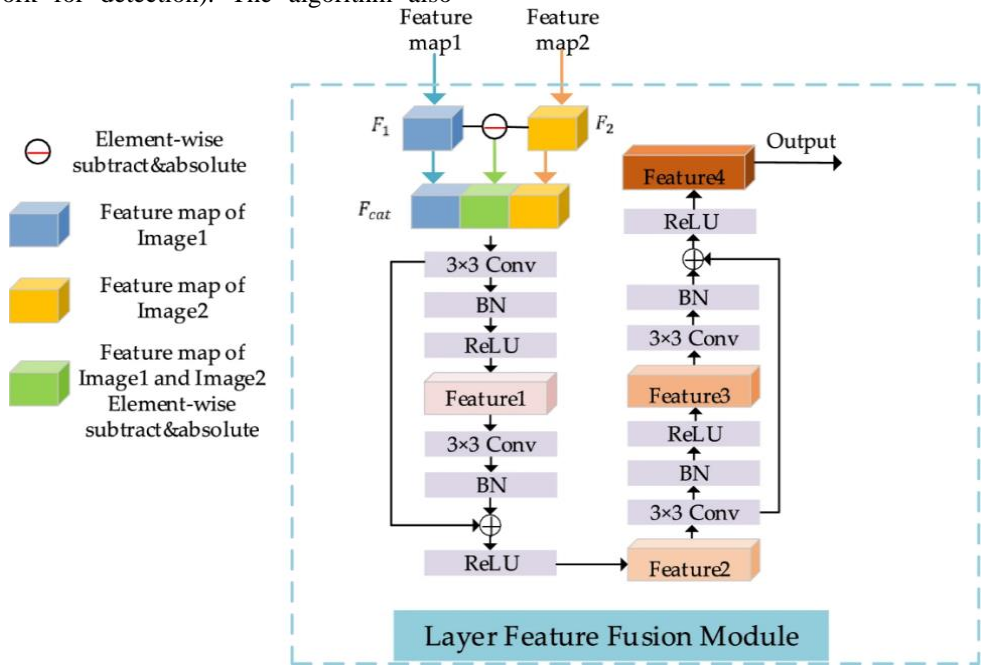

Fig. 1. Improved multi-source information fusion network based on U-net

### B. Reconfiguring the Network Architecture

The compressed path is combined with the ResNet block, using a shortcut connection. The architecture of the ResNet block is shown in Figure 2, which is a 3x3s convolutional network and jump connection. By superimposing the matrix of input and output at each level of the backbone network, the method can maintain the identity transformation when no knowledge is obtained[24].

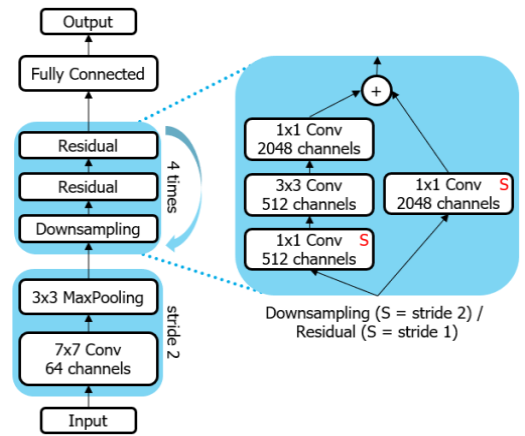

Fig. 2. ResNet block structure

## C. Information Enhancement Module

### 1) Context enhancement module

The content enhancement model is implemented with 3x3 convolution 5x3 depth separable volume. 3x3 extended convolution can accommodate more content without affecting the intrinsic properties of the image and thus have a wider perceptual domain. Multiple eigenmatrices equal to the number of channels are obtained by convolution operation for each channel. Multiple feature graphs are obtained by point-by-point convolution, which not only ensures the training accuracy, but also reduces the computational complexity [25]. The context enhancement module is shown in Figure 3 (image referenced in the multi-scale semantic network for detection). Relu is used to enhance the nonlinearity of the model, skip the output feature information and the input of CE block, and the fused feature information is used as the output of CE block to enhance the context information in multiple scales.

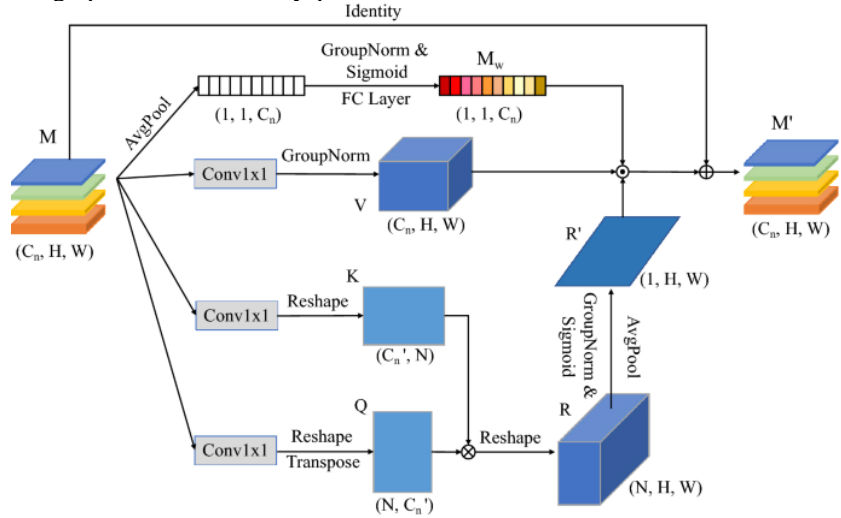

Fig. 3. Context enhancement module

### 2) Convolution pool pyramid in void space

In this paper, a cavity model based on dilatational convolution is proposed, which uses extended convolution with a certain dilatational coefficient to achieve multi-scale image description [26]. In Figure 4, four 3x3s extended convolution is used to construct the feature map. The algorithm can effectively overcome the problem of image distortion and ensure the accuracy of the image.

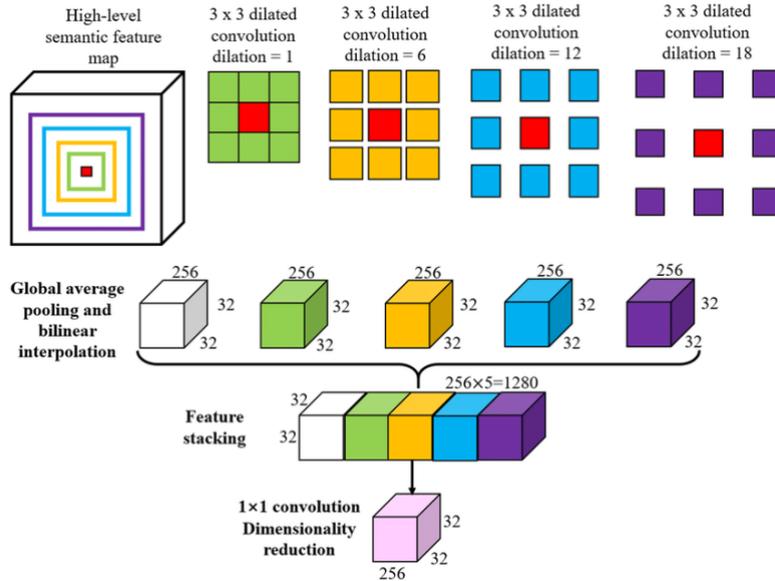

Fig. 4. Convolution pool pyramid of empty space

## D. Decoder

The decoder can significantly improve the decoding efficiency. In the process of expansion, deconvolution, and up-sampling are usually used to restore the size of the image, while deconvolution can reconstruct the image. As can be seen from Figure 5, the decoding block consists of two 1-by-1 convolution and 3-by-3 deconvolution, which can well restore the high-resolution characteristics of the image (image quoted in Pined: 3D Object Detection with Point Features).

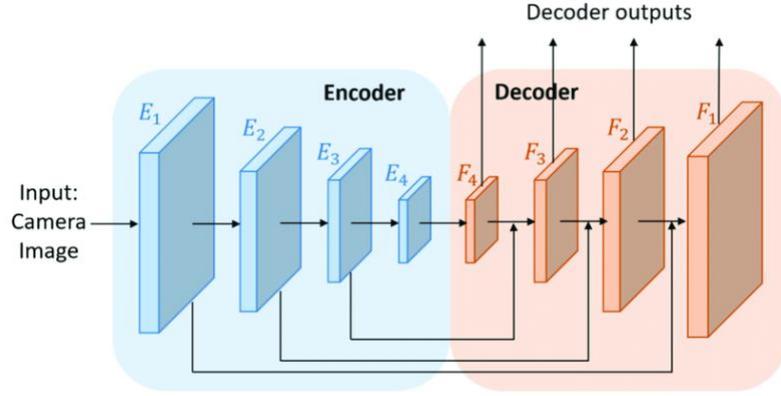

Fig. 5. Decoder module

## III. TEXTURE 3D RECONSTRUCTION ALGORITHM

### A. Basic principle of binocular reconstruction

In this project, the target image taken by two cameras of the same size is converted into distance by triangle division method. Through pixel coordinate system to camera coordinate system. The relationship between parallax and target 3D coordinates is as follows

$$\begin{cases} Z_\varepsilon = \dfrac{gS}{c} \\ X_\varepsilon = \dfrac{(x_1 - x_\varepsilon)}{g} Z_\varepsilon \\ Y_\varepsilon = \dfrac{(y_1 - y_\varepsilon)}{g} Z_\varepsilon \end{cases} \quad (1)$$

$(X_\varepsilon, Y_\varepsilon, Z_\varepsilon)$ is the coordinate of a point $Q$ on the object in the left camera coordinate system. Entity matching is all about getting a parallax map. This step is the key step of stereo vision reconstruction, and its quality will have a great impact on the accuracy of stereo vision reconstruction. The 3D shape of the object can be reproduced by the obtained parallax curve and formula (1).

### B. Image matching of multi-scale image fusion using weighted MLS filtering

$R_0$ is the image after correction, background removal, and clipping. $R_1$ is the base layer of $R_0$ after weighted least squares filtering. $R_2$ is the base layer of $R_1$ after weighted least squares filtering. $R_3$ is the base layer of $R_2$ after weighted least squares filtering. Firstly, two corrected binocular images were divided into four levels by weighted least squares filter. An algorithm based on absolute gray difference, gradient and nonparametric conversion is proposed. Guided filtering is used for cost aggregation. A multi-scale image data fusion method is constructed to obtain parallax map, and 3D modeling of road surface is reconstructed based on the parallax map. In this method, the original image is filtered by weighted least squares, and then the difference between the original image and the original image is taken as the detailed information of the image. Using this method to protect the image boundary, the detail information can be obtained from multiple levels effectively and the influence of noise can be reduced. This method takes the underlying data as input, and carries out multi-level decomposition. The calculation principle of the weighted least squares filter is expressed in formula (2), and the expression of the filtered base layer image $\sigma$ is

$$\sigma = \min_\sigma \sum_q \left\{ (\sigma_q - h_q)^2 + \eta \left[ \lambda_{x,q} (\nabla \sigma_{x,q})^2 + \lambda_{y,q} (\nabla \sigma_{y,q})^2 \right] \right\} \quad (2)$$

$h$ is for the original. $q$ represents the coordinates of a pixel, and $\nabla \sigma_{x,q}, \nabla \sigma_{y,q}$ is the gradient value of point $q$ in the $x$ and $y$ directions, respectively. $\lambda_{x,q}, \lambda_{y,q}$ represents the weights assigned by point $q$ in the $x$ and $y$ directions, respectively. $\eta$ is the parameter that balances the two terms.

### C. Multi-scale information fusion model

The matching cost function is constructed by combining absolute gray difference method, gradient method and non-parametric method. The existing matching cost algorithm only uses part of features and is easily disturbed by noise. In the case of poor texture, it is difficult for this method to reflect the correlation between features. The cost fusion technology makes the correlation between pixels more accurately through the correlation analysis of adjacent pixels. The principles for using guided filters to aggregate costs in a single scale are:

$$\begin{cases} \tilde{E}(i,c) = \dfrac{1}{N_i} \sum_{j \in W_i} L(i,j) E(j,c) \\ L(i,j) = \dfrac{1}{|\lambda|^2} \sum_{l:(i,j) \in \lambda_l} \left[ 1 + \dfrac{(W_i - \delta_l)(W_j - \delta_l)}{\varphi_l^2 + \xi} \right] \end{cases} \quad (3)$$

$\tilde{E}(i,c)$ represents the aggregation cost of pixel $i$ when parallax is $c$. $E(j,c)$ is the matching cost of pixel $j$, calculated by the matching cost function constructed based on the fusion of gray absolute difference, gradient and non-parametric transformation. Pixel $j$ is a neighborhood of pixel $i$. $N_i$ is a normalized constant. $L(i,j)$ is the kernel function of the guide filter. $\lambda$ represents the number of pixels that support window $\lambda_l$. $l$ is the center pixel of the support window. $\xi$ is the penalty factor. $W$ is the guide image. $\delta_l$ is the mean of the window pixels. $\varphi_l^2$ represents the variance of the boot image $W$ in window $\lambda_l$. The relationship between the aggregation cost after the fusion of information of different scales and the aggregation cost of a single scale is

$$\begin{bmatrix} 1+\zeta & -\zeta & 0 & 0 \\ -\zeta & 1+2\zeta & -\zeta & 0 \\ 0 & -\zeta & 1+2\zeta & -\zeta \\ 0 & 0 & -\zeta & 1+\zeta \end{bmatrix} \times \begin{bmatrix} \hat{E}^0(i^0,c^0) \\ \hat{E}^1(i^1,c^1) \\ \hat{E}^2(i^2,c^2) \\ \hat{E}^3(i^3,c^3) \end{bmatrix} = \begin{bmatrix} \tilde{E}^0(i^0,c^0) \\ \tilde{E}^1(i^1,c^1) \\ \tilde{E}^2(i^2,c^2) \\ \tilde{E}^3(i^3,c^3) \end{bmatrix} \quad (4)$$

$\hat{E}(i^s,c^s)$ is the aggregation cost of pixel $i$ (in the left image) when parallax is $c$ at scale $s$ after multi-scale information aggregation. $\tilde{E}(i^s,c^s)$ is the aggregation cost after aggregation by guided filtering.

## IV. EXPERIMENTAL RESULTS AND ANALYSIS

The paper used medical imaging data from patients with the National Institute of Cancer funded Tumor Genetic Map. In this project, 110 TCGA patients were selected as research objects, and the liquid decay reversal sequence and whole genome clustering were performed on them. 900 images were selected as training samples and 90 images were selected as test samples. The dataset is handled through the Linked Data methodology, which consolidates various data formats, a crucial aspect in academic studies[27]. This organized technique facilitates the cross-referencing of data, boosting the interoperation within datasets. The training component was repeated 300 times with a learning rate of 0.001 and 8 batchsize. It is compared with the existing classical Unet protocols such as U-net and ResU-net[28]. The network architecture adopts two layers of convolutional networks and one layer of maximum buffer layer. In the part of feature quantity coding, the pre-trained ResNet-34 is used instead of the encoder, and only 4 feature extraction modules which have not been homogenized are retained. At the same time, a fast algorithm is added to prevent gradient loss and speed up the convergence of the network. The results of the partitioning are shown in Figure 6.

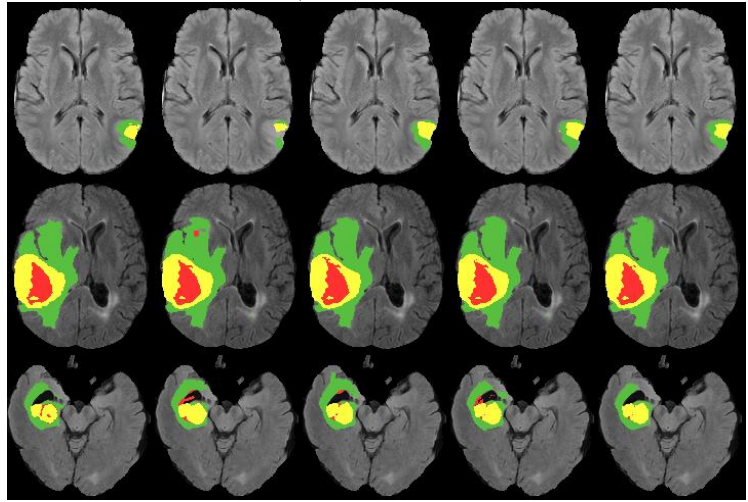

Fig. 6. Experimental results

Among them, the first column is the source of glioma cells in the brain, from left to right, U-net, ResU-net and the segmentation effect of this study model; The segmentation of Unet and ResU-net images is defective, and the processing effect is not ideal. This method can not only accurately determine the boundary of brain tumor, but also has good noise suppression ability. Compared with the traditional U-net and ResUnet, this algorithm not only improves ACC, AUC and Sen, but also improves the learning efficiency of the algorithm.

TABLE I. QUANTITATIVE EVALUATION OF EXPERIMENTAL RESULTS

| Method | U-net | Res U-net | Proposed Method |
|---|---|---|---|
| ACC | 0.9645 | 0.9649 | 0.9669 |
| AUC | 0.8142 | 0.8444 | 0.8477 |

| | Sen | 0.9879 | 0.9890 | 0.9893 |

## V. CONCLUSION

This project aims to introduce a residual-based U-Net network that connects data from upper and lower layers through skip connections and incorporates two modules—Context Encoding (CE) and Atrous Spatial Pyramid Pooling (ASPP)—to enhance contextual understanding. This enhancement is anticipated to improve the model's cognitive performance significantly. The results indicate effective edge detection and segmentation of brain gliomas, with the method efficiently integrating multiple levels of data. This novel approach offers significant potential for advancing clinical medical image segmentation on a broader scale.